# A Review of Popular Applications on Google Play – Do They Cater to Visually Impaired Users?

**Gayatri Venugopal***

Symbiosis Institute of Computer Studies and Research, Symbiosis International University, India;
gayatri.venugopal@sicsr.ac.in

## Abstract

The number of applications on online mobile application stores is increasing at a rapid rate. Smart-phones are used by a wide range of people varying in age, and also in the ability to use a smart phone. With the increasing dependency on smart-phones, the paper aims to determine whether the popular applications on Google Play, the official store for Android applications, can be used by people with vision impairment. The accessibility of the applications was tested using an external keyboard, and TalkBack, an accessibility tool developed by Google. It was found that several popular applications on the store were not designed keeping accessibility in mind. It was observed that there exists a weak positive relationship between the popularity of the application and its accessibility. A framework is proposed that can be used by developers to improve the accessibility of an application. The paper also discusses the programming aspects to be considered while developing an Android application, so that the application can be used by sighted as well as visually impaired users.

## 1. Introduction

'Disability is the inability or limitation in performing tasks, activities, and roles to levels expected in physical and social contexts'[1]. Visual impairment may refer to total blindness, partial blindness, or color blindness, wherein total blindness is a condition where the person's vision has been lost completely. A partially blind person can see images, but not clearly, and a color blind person may get confused while identifying a particular color or a combination of colors[2]. Hearing impairment is the partial or total loss of hearing, and motor impairment is the inability to perform certain physical activities associated with one or more body parts. The paper focuses on visual impairment and the smart-phone applications that can be used by people with partial or complete blindness. The paper, however, does not focus on web-based applications.

According to World Health Organization, in the year 2013, 285 million people were visually impaired across the world, and approximately 90% of the visually impaired people live in developing countries. Around 19 million children around the world are visually impaired[3]. World Health Organization, in the World Report on Disability of 2011[4], specifies lack of accessibility as a major factor that converts impairment into a disability. Lack of accessibility may refer to transport, print, infrastructure or any entity that the person is unable to make use of due to his/her impairment. With the ongoing growth of technology and its significance in our daily activities, it is essential that these smart-phones and the applications provided by developers and companies cater to the needs of people with impairments. All users should not be compelled to use the same method of interaction with the application, without considering their specific needs[5]. The market today is booming with mobile operating systems from various companies, the front runners being, Android by Google, and iOS by Apple. According to a report by IDC, Android dominates the smart-phone market share as of February 2014[6]. Owing to its open source nature, Android has a vast developer community that develops

*Author for correspondence

applications and uploads them on the official market, called Google Play. Android phones are extremely popular in developing countries such as India as a wide variety of options with respect to price are available to the buyer. Since Android phones can be found in a huge number in developing countries, and these countries cover the majority of the percentage of people with vision impairment across the world, the objective of the paper is to explore the popular applications available on Google Play and to identify whether they can be used by visually impaired users. The paper also aims to determine whether there is any relationship between the popularity of the app and its accessibility.

## 2. Efforts to Facilitate Accessibility

Substantial amount of work has been done in this domain leading to several suggestions from researchers. Kientz et al[7], developed an application for locating lost items using a tagging system. In order to reduce the effort associated with text entry, the user's voice is used as the input. The authors suggest that an auditory interface could be used to adapt to devices with small screens. Batusek and Kopecek[8] recommend the development of an easily customizable system, wherein the user should be able to retrieve the required content quickly. In case of a voice recognition system, the authors state that the developer should take into consideration a scenario where one command may be spoken by the user in several ways. de Sa and Carrico[9] emphasize on user interfaces with high contrast ratio, and on the size of the elements on the screen. A partially blind person should not be unable to use the application due to the small size of buttons or other user interface elements. Provision of a set of preferences to choose from, ease of use with one hand, and availability of alternative descriptions of the elements on the screen are some of the solutions suggested by Kukulska-Hulme[10]. The User Centred System Design (UCSD) framework was proposed by Norman and Draper[11] where the user's requirements play a key role in the design and development of a software product. Gulliksen et al[12], describe the inclusion of UCSD in various phases of the system life-cycle. It is included in the planning phase, wherein the plan for the design is determined; in the analysis phase, the requirements of the users are ascertained; in the design phase, a detailed design is created; a continuous evaluation of the software is performed in order to identify usability issues, and the feedback thus obtained is used as an input for the planning phase, thus iterating the steps specified above. After successful results, the software is implemented and monitored with respect to its usability. Throughout the process, there should be a continuous focus on the users and their requirements. Newell and Gregor[13] suggest the term 'User Sensitive Inclusive Design' wherein people with impairments are made part of the design and development process and the results thus obtained are shared with other researchers, designers and developers. Since the intensity of impairment may vary from person to person, his/her way of interacting with a smart-phone may also vary, which means the difficulty faced by people with visual impairment may vary depending on their usage[14]. Therefore while considering users with impairments, the difference in usage patterns should also be considered. Sometimes involving users in all the stages may not be a feasible solution. In such cases, designers and developers can make use of simulators to understand the accessibility problems and to design, develop and test the application accordingly[15].

## 3. Methodology

Google Play (initially Android Market), the official store for purchasing or freely downloading Android apps, has over a million applications, consisting of both, free as well as paid applications[16]. The user base of Android applications consists of users from various walks of life, ranging from students and working people to senior citizens. Thus applications should be developed keeping in mind a wide variety of users. With the launch of Android One[17], more users will have access to budget phones that run on an Android platform. The author tested 53 applications uploaded on Google Play, for parameters such as usefulness with accessibility service (TalkBack) enabled, navigation through an external keyboard, and the ability to change text size and color. Popular applications from books, communication, education, finance, health, media, news, productivity, shopping, social, tools and weather categories were tested. The applications were downloaded on Aakash 2 tablets and Micromax A74 smart-phone. The popularity of applications was based on the reviews and the number of downloads of the application, that is displayed on the application's page on the store. Table 1 lists the names and details of the applications that were tested, and Table 2 lists the observations made with respect to navigation capability using TalkBack accessibility service and/or an external keyboard, and user's ability to customize the application's text size and colors. The details in Table 1 have been retrieved from Google Play[18].

**Table 1.** List of applications and their respective details

| Name of the Application | Description | Current Version | OS Version Support |
|---|---|---|---|
| Uber | Request a ride using the Uber app and get picked up within minutes. | 3.11.1 | >=4.0.3 |
| Zomato – Restaurant Finder | Search for restaurants, view menus and user reviews. | 5.4.2 | Varies with device |
| Yatra.com | Travel company | 4.0.6 | >=2.3 |
| MakeMyTrip | Travel company | 3.2.2 | >=2.3.3 |
| Maps | Makes navigation easier. | 6.14.5 | Varies with device |
| Ccleaner | Removes unnecessary data, reclaims space, and helps monitor the system and browse safely. | v1.02.20 | >=4.0 |
| AirDroid | Helps manage Android devices on the web, all over the air. | 2.1.0 | >=2.2 |
| Calculator Plus Free | Allows the user to perform calculations and stores the data, which can be reviewed by the user later. | 4.8.0 | >=2.3.3 |
| Fast Notepad | Text editor. | 1.4.4 | >=1.6 |
| Google Gesture Search | User can use gesture to perform operations. | Varies with device | Varies with device |
| Facebook | See what friends are up to. Share updates, photos and videos, get notified when friends like and comment on your posts, text, chat and have group conversations, play games and use your favorite apps. | Varies with device | Varies with device |
| Twitter | Twitter is a free app that lets you connect with people, express yourself, and discover more about all the things you love. | Varies with device | Varies with device |
| LinkedIn | Build your personal brand, make connections, stay informed with personalized news. | Varies with device | Varies with device |
| WordPress | Write, edit, and publish posts to your site and check status. | Varies with device | Varies with device |
| Flipkart | Users may choose from the massive selection of original products in Fashion, Electronics, Books, Mobiles and other categories. The app also provides timely alerts on great deals with substantial discounts. | 3 | >=2.3 |
| Amazon | The Amazon app lets you shop and manage your Amazon orders from anywhere. Browse and shop by department, compare prices, read reviews, share products with friends, check out Gold Box Deals, make purchases, and check the status of your orders. | 5.0.1 | Varies with device |
| Adobe Reader | Using Adobe reader, the user may work with PDF documents on their Android tablets or phones. It allows the user to easily access, manage, and share a wide variety of PDF types. | 11.5.0.1 | >=2.3.3 |
| ES File Explorer File Manager | ES File Explorer File Manager is a free, full-featured file manager. It manages files using Multiple Select, Cut/Copy/Paste, Move, Create, Delete, Rename, Search, Share, Send, Hide, Create Shortcut, and Bookmark. | 3.2.1 | Varies with device |
| Microsoft Office Mobile | User can access, view and edit Microsoft Word, Microsoft Excel and Microsoft PowerPoint documents from virtually anywhere. | 15.0.2720.2000 | >=4.0 |
| Dropbox | Dropbox is a free service that lets the user bring all photos, docs, and videos anywhere. | Varies with device | Varies with device |
| Evernote | Evernote lets the user create and manage notes efficiently. | Varies with device | Varies with device |
| Google Drive | All the files on Google Drive can be accessed easily on the Android device. | Varies with device | Varies with device |
| OneDrive | User can easily store and share photos, videos, documents, and more. When the user uploads files from an Android device to OneDrive, he/she can get to them on a PC, Mac, tablet, or phone. With OneDrive for Android, the user can easily get to, manage, and share files on the go. | 2.7.0 | Varies with device |
| CNN Breaking US and World News | Helps the user stay informed with the latest headlines and original stories from around the globe. | 2.5 | >=2.3.3 |
| Google Play Newsstand | Breaking news and in-depth articles featuring audio, video and more. | 3.2.1 | >=2.3 |
| LinkedIn Pulse | Pulse by LinkedIn is the professional news app tailored to you. Pulse allows you to customize your news reading experience, easily explore compelling professional content, and share stories to your favorite social networks. | 4.1.12 | >=2.2 |
| The Guardian | App to read contents of The Guardian newspaper. | 2.7.10 | Varies with device |
| MX Player | Video player with support for hardware acceleration, multi-core decoding, pinch to zoom, zoom and pan, subtitle gestures. | 1.7.31 | >=2.1 |
| You Tube | User can view videos available on YouTube. | 5.10.1.5 | Varies with device |
| Google Play Music | User can listen to unlimited songs, create custom radio from any song, artist or album, enjoy radio without skip limits, get smart recommendations based on his/her tastes. | 5.6.1623P.1416251 | Varies with device |
| SoundHound | SoundHound recognizes music playing around you. Tap the SoundHound button to instantly identify songs and see lyrics, share, buy or explore more from artists you know and love or have just discovered. | 6.2.0 | Varies with device |
| Nike Training Club | The user can choose individual workouts, or select a targeted, structured four-week program. Workouts can be customized according to the user's preferences. | 3.1 | >=3.1 |
| Calorie Counter – MyFitnessPal | Consists of a database of over 3,000,000 food items. Helps to keep track of the calorie intake by the user. | 3.4 | Varies with device |
| Runtastic Pedometer | Whether walking, hiking, or backpacking for pure enjoyment, to improve the user's general health and fitness, or as part of a specific diet and weight loss routine, the Runtastic Pedometer counts every step taken. | 1.5 | Varies with device |
| Water Your Body | This app reminds the user to drink water every day and tracks your water drinking. | 3.031 | >=2.3 |
| Google Finance | It synchronizes with the user's Google Finance portfolios, allows quick access to charts and lets him/her view the latest market and company news. | 2.2.7 | >=1.5 |
| Bloomberg for Smartphone | Displays business/finance news, market data and stock tracking tools. | 1.2.4.100 | >=1.6 |
| StockSpy | This app which simplifies tracking stocks, stock market, real-time quotes, charts, news, links and stats for stocks around the globe. | 2.3 | >=2.2 |
| Expense Manager | Simple, intuitive, stable and feature-rich app that helps the user to manage the expenditures and budgets. | 2.1.2 | Varies with device |
| Speak English | The user can improve his/her pronunciation and speaking skills. | 2.3 | >=2.3.3 |
| Coursera | Coursera connects students, professionals, and lifelong learners everywhere with free online courses from over 100 top-tier global universities and institutions. | 1.2.1 | >=4.0 |
| NASA Archives | An app about NASA and the history of human space flight. | 1.2 | >=2.2 |
| IELTS Word Power | Vocabulary practice app for IELTS test takers. | 1.0 | >=1.5 |
| Skype | Allows instant message, voice or video call for free. | 5.0.0.5727 | Varies with device |
| Viber | More than 400 million Viber users text, call, and send photo and video messages worldwide over Wi-Fi or 3G - for free. | 5.0.1.36 | >=2.3 |
| Gmail | Mailing app by Google. | 4.0.4-255454 | Varies with device |
| Yahoo Mail | Mailing app by Yahoo. | 4.6.2 | >=2.3 |
| Oxford Dictionary | Offers a comprehensive coverage of English from around the world. | 4.3.106 | >=2.2 |
| Google Play Books | Choose from millions of books on Google Play including new releases, New York Times bestsellers, textbooks and free classics. | 3.1.49 | Varies with device |
| Audible for Android | The user can download books and listen to them on the go. | 1.5.7 | Varies with device |
| Wikipedia | Wikipedia is the free encyclopedia containing more than 32 million articles in 280 languages, and is the most comprehensive and widely used reference work humans have ever compiled. | 2.0-r-2014-08-13 | >=2.3.3 |
| Yahoo Weather | Gives the latest weather conditions. | 1.2 | >=2.3 |
| AccuWeather | Gives the latest weather conditions. | 3.3.1.1 | Varies with device |

**Table 2.** Observations made with respect to the accessibility of the applications

| Name of the Application | Can the Application be Operated using TalkBack and Explore by Touch? | Can the Application be Operated using an External Keyboard? | Can Text Size be Changed? | Can Font/ Background Color be Changed? | Remarks |
|---|---|---|---|---|---|
| Uber | Gave speech feedback as Button[37], unlabelled. Buttons with images do not have content Description. Items on the map are not read. Pane displaying the cars does not give any voice feedback. | Navigation is possible but no feedback is given to the user. | No | No | The app is dependent majorly on touch events to set the pickup location and to select a car. |
| Zomato – Restaurant Finder | Buttons with images do not have content Description. | No issue detected. | No | No | The help screen that appears when the app is launched for the first time does not give any speech feedback. |
| Yatra.com | Gave speech feedback as Button. Images indicating adult and children do not give appropriate feedback. Buttons with images do not have contentDescription. | No issue detected. | No | No | The help screen that appears when the app is launched for the first time does not give any speech feedback. |
| MakeMyTrip | Buttons with images do not have contentDescription. | No issue detected. | No | No | Works well on the whole as there are very few buttons without text. |
| Maps | The text on the map was not read out for the user. | Unable to shift focus to tabs, from the map. | Yes | No | Works well when using functionality such as search, get directions. The user cannot decide the amount by which the font size should be increased. The font size for the text on the map is increased, by that of on buttons and labels remains constant. |
| Ccleaner | Images did not have content description. | Unable to switch between tabs. | No | No | Appropriate feedback was given for all the views except images that appear in the list of applications. |
| AirDroid | Yes | Incorrect flow of navigation. | No | No | Switches are placed at a distance from their label, and are not enabled/disabled when the user selects the label. This could occur because the label and checkbox are not associated with each other. While using the arrow keys on the keyboard, if the end of the row is reached, the focus should shift to the next line. Instead, the focus shifts to the next tab. |
| Calculator Plus Free | Gave speech feedback as Button [8], unlabeled. Buttons with images do not have contentDescription. | No | No | No | The app cannot be used with the help of Talk Back alone as the user does not get the appropriate feedback about the button that was pressed by him/her. The app cannot be used using a keyboard as it does not detect button selection through the keyboard. |
| Fast Notepad | Gave speech feedback as Button [60], unlabeled. Buttons with images do not have contentDescription. | Yes | Yes | Yes | Font color can be changed to either black or white. |
| Google Gesture Search | Yes | NA | No | No | Checkboxes are placed at a distance from their label, and are not enabled/disabled when the user selects the label. This could occur because the label and checkbox are not associated with each other. |
| Facebook | Yes | Unable to shift focus to the bar at the top that has options to search and view contacts. | No | No, but the existing color combination is good, as it consists of a white background and black text. | |
| Twitter | Yes | Unable to shift focus to the bar at the top that has options to create a new post, search people etc. | Yes | No, but the existing color combination is good, as it consists of a white background and black text. | The app is accessible. |
| LinkedIn | Gave speech feedback as Image[32], unlabeled. Images do not have contentDescription. | Unable to select one topic from the list of contents. | No | No | |
| WordPress | Gave speech feedback as Image[4], unlabeled. Images do not have contentDescription. | No issue detected. | No | No | |
| Flipkart | Gave speech feedback as Image[62], unlabeled. Images do not have contentDescription. | Unable to browse through the list of items on the screen, thus unable to select an item. | No | No | |
| Amazon | Gave speech feedback as Button[67], unlabeled. Buttons with images do not have contentDescription. | No issue detected. | No | No | |
| Adobe Reader | Yes | Can navigate between pages without any problem, but the text to speech module does not start reading the content until the user touches the screen. | No, but text can be enlarged by zooming in using the pinch gesture. | No | |
| ES File Explorer File Manager | Gave speech feedback as Button[28], unlabeled. Buttons with images do not have contentDescription. | Yes | No | No | |
| Microsoft Office Mobile | Gave speech feedback as Button[22], unlabeled at some places. Some buttons with images do not have contentDescription. No feedback when a cell is selected in Excel. | Yes | No, but text can be enlarged by zooming in using the pinch gesture. | No | |
| Dropbox | No feedback for image button, implying no content Description is associated with the button. | Yes | No | No | |
| Evernote | Screen with images alone that describes the features of Evernote does not give any voice feedback. Gave speech feedback as Button[77], unlabeled at some places. Some buttons with images do not have content Description. | Unable to add note using keyboard. Focus not received on the add button. | No | No | |
| Google Drive | Yes | Yes | No, but text can be enlarged by zooming in using the pinch gesture. | No, but the user may change the font color of the content in the editing mode. | |
| OneDrive | Yes | Options at the top disappear when the note is displayed on the screen, thus making it difficult to switch back to the options without selecting the area using touch. | No | No | |
| CNN Breaking US and World News | No speech feedback for content in the article. | No speech feedback for content in the article. | No | No | |
| Google Play Newsstand | Yes | Unable to select an article to view the complete content. | Yes | No | |
| Linked In Pulse | Gave speech feedback as Button[8], unlabeled. Buttons with images do not have contentDescription. | No speech feedback for content. | No | No | |
| The Guardian | No feedback for the content in the article. | No issue detected, text to speech feedback for content is possible using keyboard. | Yes | No | |
| MX Player | Yes | No issue detected. | Yes | Yes | The app is accessible. |
| You Tube | Yes | Unable to select a video in the list. | Yes | Yes | |
| Google Play Music | Yes | No issue detected. | No | No | |
| Sound Hound | Yes | No option to record audio without touch. No other issue detected. | No | No | |
| Nike Training Club | No feedback for image button, implying no contentDescription is associated with the button. | No issue detected. | No | No | |
| Calorie Counter – My Fitness Pal | Yes | Unable to shift focus to various options on the screen. | No | No | |
| Runtastic Pedometer | No content description for the lock button, which is used to stop or pause the counting of steps. No content description for radio buttons that are displayed in the form of images. | Unable to shift focus to the number of steps taken, and duration labels. Unable to shift focus to the labels that describe the environment in which the user went for a walk. | No | No | Since the app will be used while walking, navigation problem using the keyboard may not be a major issue. |
| Water Your Body | Yes | Unable to shift focus to various options on the screen. | No | No | |
| Google Finance | No content description for image displaying charts. No speech feedback for content in the news section. | No other issue detected. | No | No | |
| Bloomberg for Smartphone | Yes | No issue detected. | No | No | |
| StockSpy | No speech feedback for the stock figures in the list. | Unable to shift focus to the stock figures in the list. | No | No | |
| Expense Manager | Yes | No issue detected. | No | Yes | |
| Speak English | Gave speech feedback as Button [56], unlabeled. | Unable to navigate using the keyboard. | No | No | |
| Coursera | Yes | No issue detected. | No | No | |
| NASA Archives | Yes | No issue detected. | No | No | |
| IELTS Word Power | No content description associated with buttons. | No other issue detected. | No | No | |
| Skype | Yes | No issue detected. | No | No | |
| Viber | Yes | No issue detected. | No | No | |
| Gmail | Yes | No speech feedback when an item in the list of mails receives focus. | Yes | No | |
| Yahoo Mail | Gave speech feedback as Button [39], unlabeled. | No speech feedback when an item in the list of mails receives focus. | Yes | No | |
| Oxford Dictionary | No speech feedback for content/ meaning of the word. | No other issue detected. | Yes | No | |
| Google Play Books | Yes | Unable to shift focus to the content area of the screen. | Yes | Yes | |
| Audible for Android | Yes | Unable to download book. | No | No | |
| Wikipedia | Yes | Unable to navigate to settings option on the right. | Yes | Yes | |
| Yahoo Weather | Gave speech feedback as Button[79], unlabelled. Buttons with images do not have contentDescription. | No issue detected. | No | No | |
| AccuWeather | Gave speech feedback such as "Image[19] unlabelled", indicating that contentDescription has not been set. Tabs did not give correct feedback. | No issue detected. | No | No | |

## 4. Results and Analysis

The objective was to determine whether there is any relationship between the accessibility of the application and its popularity which is measured by the number of downloads and the average user ratings. The model given below was devised to quantify the accessibility of an application. The following variables were considered:
a: application is accessible using TalkBack
b: application is accessible using external keyboard
c: user can change the text size of the content
d: user can change the text/background color of the content,
Accessibility coefficient of an application X, $(X_{accessibility}) = f(a, b, c, d)$, where

$$f(a,b,c,d) = \begin{cases} 1 & \text{if } a.b.c.d = 1 \\ 0.5 & \text{if } (a.b)+(c.d) = 1 \\ 0 & \text{if } (a.b)+(c.d) = 0 \end{cases}$$

The model was developed based on the fact that for a blind user, the text color and size are insignificant, thus the application is accessible to him/her if there is appropriate speech feedback when using the touch screen or while navigating using an external keyboard. On the other hand, for a partially blind user, the text size and colors play an important role in the usage of the app. Therefore in this case the application is accessible if the user is provided with the option to modify the text size and text/background colors.

The accessibility coefficients for the 53 applications were calculated, and compared with their corresponding popularity. The popularity is the sum of the average user rating and the number of downloads (scaled down to a smaller number) of each application. Here, popularity is denoted by g(r,dl), which is calculated as r + dl/100000, where r refers to the average user rating and dl refers to the number of downloads. Table 3 denotes the popularity and accessibility coefficients of the applications calculated using the specified model.

## 5. Discussion

As can be seen, many applications discussed here follow the cell of exclusion19, wherein special needs of users are not taken into consideration. In order to improve the accessibility of an application, the following points should be considered by developers in order to create an application that supports inclusion:

1. The user must be able to change the default font size, or he/she must be able to zoom in our out using simple gestures.
2. The user must be given an option to choose from a list of text and background colors.
3. Content descriptions in Android applications are used by screen readers and other accessibility services to describe to the user, the elements present on the screen. It should be a string that identifies the element with which it is associated. As can be seen, it was found that majority of the applications did not have appropriate content description text associated with the elements. In Android, the content description can be specified in the XML file using the android:contentDescription attribute, or in the java code using the setContentDescription method.
4. Keyboards are often used with touch screen devices to interact with applications; hence it is important that the focus is shifted from one element to the other in the correct sequence, when the user navigates between the elements on the screen using the tab and/or arrow keys.
5. If the user interacts with the application using an external keyboard, the provision to navigate between screens, items in a list, spinner etc. should be imple-mented and tested using the corresponding keys on the keyboard.

**Table 3.** Popularity and accessibility coefficients of the applications

| Name of the Application | Average Rating r | Downloads dl | Popularity P = g(r, dl) | a | b | c | d | $X_{accessibility}$ = f(a, b, c, d) |
|---|---|---|---|---|---|---|---|---|
| Uber | 4.2 | 1000000 | 14.2 | 0 | 0 | 0 | 0 | 0 |
| Zomato – Restaurant Finder | 4.3 | 1000000 | 14.3 | 0 | 1 | 0 | 0 | 0 |
| Yatra.com | 4 | 100000 | 5 | 0 | 1 | 0 | 0 | 0 |
| MakeMyTrip | 4.1 | 1000000 | 14.1 | 0 | 1 | 0 | 0 | 0 |
| Maps | 4.3 | 1000000000 | 10004.3 | 0 | 0 | 1 | 0 | 0 |
| Ccleaner | 4.4 | 1000000 | 14.4 | 0 | 0 | 0 | 0 | 0 |
| AirDroid | 4.6 | 10000000 | 104.6 | 1 | 0 | 0 | 0 | 0 |
| Calculator Plus Free | 4.4 | 1000000 | 14.4 | 0 | 0 | 0 | 0 | 0 |
| Fast Notepad | 4.4 | 5000000 | 54.4 | 0 | 1 | 1 | 1 | 0.5 |
| Google Gesture Search | 4.3 | 1000000 | 14.3 | 1 | 1 | 0 | 0 | 0.5 |
| Facebook | 4 | 500000000 | 5004 | 1 | 0 | 0 | 0 | 0 |
| Twitter | 4.1 | 100000000 | 1004.1 | 1 | 0 | 1 | 0 | 0 |
| LinkedIn | 4.2 | 10000000 | 104.2 | 0 | 0 | 0 | 0 | 0 |
| WordPress | 4.2 | 1000000 | 14.2 | 0 | 1 | 0 | 0 | 0 |
| Flipkart | 4.3 | 5000000 | 54.3 | 0 | 0 | 0 | 0 | 0 |
| Amazon | 4.4 | 10000000 | 104.4 | 0 | 1 | 0 | 0 | 0 |
| Adobe Reader | 4.3 | 100000000 | 1004.3 | 1 | 0 | 1 | 0 | 0 |
| ES File Explorer File Manager | 4.6 | 50000000 | 504.6 | 0 | 1 | 0 | 0 | 0 |
| Microsoft Office Mobile | 4 | 5000000 | 54 | 0 | 1 | 1 | 0 | 0 |
| Dropbox | 4.5 | 100000000 | 1004.5 | 0 | 1 | 0 | 0 | 0 |
| Evernote | 4.6 | 50000000 | 504.6 | 0 | 0 | 0 | 0 | 0 |
| Google Drive | 4.4 | 100000000 | 1004.4 | 1 | 1 | 1 | 1 | 1 |
| OneDrive | 4.3 | 5000000 | 54.3 | 1 | 0 | 0 | 0 | 0 |
| CNN Breaking US and World News | 3.8 | 10000000 | 103.8 | 0 | 0 | 0 | 0 | 0 |
| Google Play Newsstand | 3.8 | 100000000 | 1003.8 | 1 | 0 | 1 | 0 | 0 |
| LinkedIn Pulse | 4.3 | 10000000 | 104.3 | 0 | 0 | 0 | 0 | 0 |
| The Guardian | 4 | 1000000 | 14 | 0 | 1 | 1 | 0 | 0 |
| MX Player | 4.5 | 100000000 | 1004.5 | 1 | 1 | 1 | 1 | 1 |
| You Tube | 4.1 | 1000000000 | 10004.1 | 1 | 0 | 1 | 1 | 0.5 |
| Google Play Music | 4 | 500000000 | 5004 | 1 | 1 | 0 | 0 | 0.5 |
| SoundHound | 4.3 | 50000000 | 504.3 | 1 | 0 | 0 | 0 | 0 |
| Nike Training Club | 4.2 | 1000000 | 14.2 | 0 | 1 | 0 | 0 | 0 |
| Calorie Counter – MyFitnessPal | 4.7 | 10000000 | 104.7 | 1 | 0 | 0 | 0 | 0 |
| Runtastic Pedometer | 4.2 | 1000000 | 14.2 | 0 | 0 | 0 | 0 | 0 |
| Water Your Body | 4.4 | 1000000 | 14.4 | 1 | 0 | 0 | 0 | 0 |
| Google Finance | 3.3 | 1000000 | 13.3 | 0 | 1 | 0 | 0 | 0 |
| Bloomberg for Smartphone | 4.1 | 1000000 | 14.1 | 1 | 1 | 0 | 0 | 0.5 |
| StockSpy | 4.4 | 100000 | 5.4 | 0 | 0 | 0 | 0 | 0 |
| Expense Manager | 4.3 | 1000000 | 14.3 | 1 | 1 | 0 | 1 | 0.5 |
| Speak English | 4 | 1000000 | 14 | 0 | 0 | 0 | 0 | 0 |
| Coursera | 4.3 | 500000 | 9.3 | 1 | 1 | 0 | 0 | 0.5 |
| NASA Archives | 4.3 | 10000 | 4.4 | 1 | 1 | 0 | 0 | 0.5 |
| IELTS Word Power | 3.9 | 500000 | 8.9 | 0 | 1 | 0 | 0 | 0 |
| Skype | 4.1 | 100000000 | 1004.1 | 1 | 1 | 0 | 0 | 0.5 |
| Viber | 4.3 | 100000000 | 1004.3 | 1 | 1 | 0 | 0 | 0.5 |
| Gmail | 4.3 | 1000000000 | 10004.3 | 1 | 0 | 1 | 0 | 0 |
| Yahoo Mail | 4.2 | 100000000 | 1000 | 0 | 0 | 1 | 0 | 0 |
| Oxford Dictionary | 3.9 | 10000000 | 103.9 | 0 | 1 | 1 | 0 | 0 |
| Google Play Books | 3.8 | 500000000 | 5003.8 | 1 | 0 | 1 | 1 | 0.5 |
| Audible for Android | 4.2 | 10000000 | 104.2 | 1 | 0 | 0 | 0 | 0 |
| Wikipedia | 4.3 | 10000000 | 100 | 1 | 0 | 1 | 1 | 0.5 |
| Yahoo Weather | 4.4 | 10000000 | 104.4 | 0 | 1 | 0 | 0 | 0 |
| AccuWeather | 4.3 | 10000000 | 104.3 | 0 | 1 | 0 | 0 | 0 |

6. Navigation between screens and elements should not be a complex process for the user if he/she is interacting with the application using the smart-phone's touch interface. Since blind users have different preferences for certain gestures as compared to sighted users[20], these gestures should be identified and implemented for navigation.

Based on these points, the author designed a generic model for improving the accessibility of a mobile application, as shown in Figure 1. While developing an application, the developer follows the requirements stated by the client. In order to make it accessible, a problem identification phase should be added to determine the barriers that people with impairments may face while using the application. According to the model, one disability should be considered at a time, and features to overcome the disability should be determined. The problems, that is, the limitations of the application with respect to one type of impairment, and their corresponding solution/s form a block called the ability structure. The designer/developer may add as many problems as observed. Each disability that needs to be catered to is associated with one ability structure. Since two or more ability structures may share one or more solutions, the solution section is not a part of any one structure. The solution/s to the disability that is, the features to be added, are placed in a separate section, the references to which are given in the ability structure. If a problem can be solved by a combination of two (or more) solutions, the identifiers of the solutions are specified, separated by a Boolean and operator (.). Similarly, if either of the solutions can solve the problem, the ids of the solutions are separated by a Boolean OR operator (+). The ability structures may be stacked on top of each other based on the priority, wherein the structure with the lowest priority should be placed at the bottom. Structures having equal priorities should be placed at the same level. It is up to the manager, designer, developer, and/or other stakeholders to determine the priority. For example, a problem faced by a visually impaired user could be a disability that should be catered to. One of the problems could be the inability to view small text. The solution could be to give an option to the user to select the text size, or to provide zoom in and zoom out features.

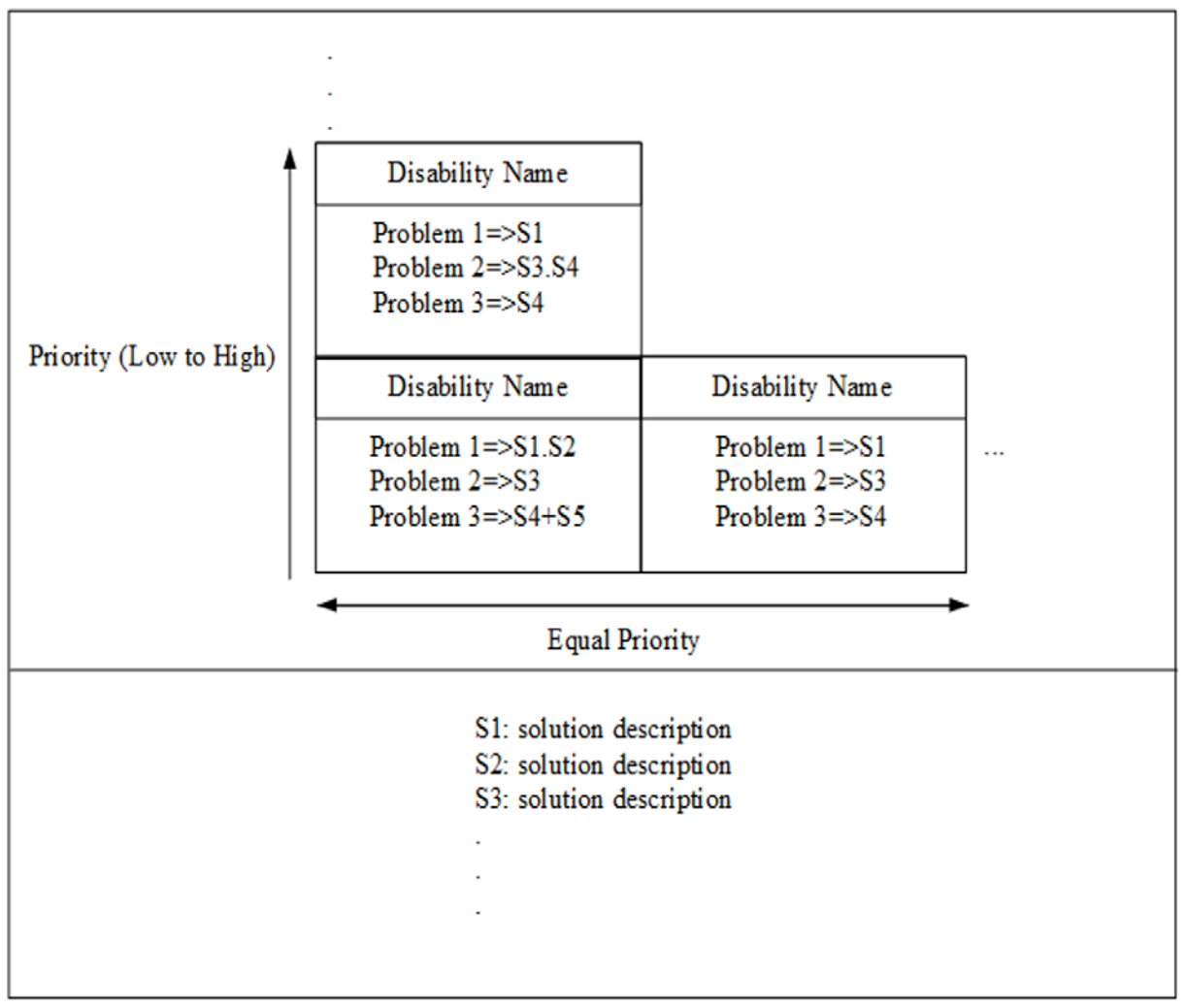

**Figure 1.** Framework for developing an accessible application.

## 6. Conclusion

The author believes that inclusive design is the need of the hour. A user's impairment should not restrain him/her from using a smart-phone. This issue can be resolved if developers and users have an interactive session, wherein the challenges faced by the user are put forth to the developer. Clients should ask for accessible features in the application, and individual developers should also ensure that the applications uploaded by them on Google Play or any other site, cater to people with various types of impairment. It is not necessary to mix the accessibility features with the other features of the application. The application could have an accessibility mode, which can be switched on or off based on the user's requirements. Software companies could also hire accessibility consultants in order to get a better understanding of the constraints that could be faced by various users. The study focused on visual impairment; more work can be done in this field by studying the problems faced by users with other types of impairment.

## Appendix: Survey Questions

1. Please enter your location (country).
2. For how many years have you been working as an Android app developer?
3. In what field/domain are you currently working?
4. What according to you is an accessible app?
5. While developing an Android app for the client, how often do you create the app keeping accessibility in mind?
   Never Rarely Frequently Always
6. What percentage of your clients asks for the inclusion of accessible features in the app?
   Less than 10% 10% to 25% 26% to 50%
   51% to 75% More than 75%
7. If you are employed with a firm, does the firm work with an accessibility consultant during the design and development of an app? If you are self-employed, do you work with an accessibility consultant to design and develop your app?
   Yes No
8. Select the factors that prevent you from making an app accessible.
   Clients do not require accessibility.
   The user base for accessible apps is small.
   Accessibility consultation is expensive.
   Unaware about accessibility testing.
   Unaware about accessibility.